\documentclass[12pt]{article}
\usepackage{amsmath}
\usepackage{amssymb}
\usepackage{multirow}
\usepackage[mathscr]{eucal}
\usepackage[usenames]{color}
\usepackage[latin5]{inputenc}
\usepackage{url}
\usepackage{graphicx}
\usepackage{epstopdf}
\usepackage{subfigure}

\newtheorem{prop}{Proposition}

\newtheorem{rmk}{Remark}

\numberwithin{equation}{section}
\numberwithin{thm}{section}
\numberwithin{lemma}{section}
\numberwithin{prop}{section}
\numberwithin{cor}{section}
\numberwithin{rmk}{section}
\numberwithin{defn}{section}

\newcommand{\semi}{\subset \hskip -4mm +}

\definecolor{darkolivegreen}{rgb}{0.333333, 0.419608, 0.1843140}

\newcommand{\dx}{\partial_x}
\newcommand{\dy}{\partial_y}

\newcommand{\dt}{\partial_t}
\newcommand{\du}{\partial_u}
\newcommand{\dv}{\partial_v}

\newcommand{\df}{\partial_{\phi}}
\newcommand{\dvf}{\partial_{\varphi}}

{

\begin{document}

\title{\Large Benney--Roskes/Zakharov--Rubenchik System: \\ Lie Symmetries and Exact Solutions\thanks{This work was supported by Scientific Research Projects Department of Istanbul Technical University, Project Number: TYL-2021-42942.} }

\author{Şeyma Gönül\thanks{Corresponding author, e-mail: gonul20@itu.edu.tr, seymaagonul@gmail.com} \quad  Cihangir Özemir\thanks{e-mail: ozemir@itu.edu.tr}\\
			\small Department of Mathematics, Faculty of Science and Letters,\\
			\small Istanbul Technical University, 34469 Istanbul,
			Turkey }
		
\date{}
		
\maketitle
		
\begin{abstract}
We investigate Lie symmetry algebra of the Benney--Roskes/Zakharov--Rubenchik systems. The invariance algebra turns out to be infinite-dimensional.
We also find several exact solutions of periodic, line-soliton and stationary types.

\vspace{1cm}
\end{abstract}

\section{Introduction}
The aim of this article is to investigate the Lie symmetry algebra of the Benney--Roskes/Zakharov--Rubenchik (BR/ZR) system and present exact solutions. We find that the symmetry algebra is an infinite dimensional Lie algebra. We succeeded in finding solutions in the forms of a line soliton, a lump type stationary solution, and periodic solutions of elliptic, hyperbolic and trigonometric type.
		
The current work was inspired by the papers \cite{Ponce2005},  \cite{Luong2018} and  \cite{Quintero2020}. Let us consider the Benney--Roskes (BR) system of \cite{Benney1969}, also appearing in  \cite{Ponce2005},
		\begin{equation}\label{BRor}
			\begin{aligned}
				&\frac{\partial b_1}{\partial T}-i \frac{\epsilon}{2}\Big\{\omega_{11}\frac{\partial^2 b_1}{\partial X^2}+\omega_{22}\frac{\partial^2 b_1}{\partial Z^2}\Big\}+i\epsilon b_1 \Big\{k\frac{\partial a_0}{\partial X}+\frac{1}{2g\omega}(g^2k^2-\omega^4)b_0\Big\}+i\epsilon |b_1|^2 b_1=0,  \\
				&\frac{\partial a_0}{\partial T}-c_g \frac{\partial a_0}{\partial X}+g b_0+\frac{1}{\omega^2} (g^2k^2-\omega^4) |b_1|^2=0,   \\
				&\frac{\partial b_0}{\partial T}-c_g \frac{\partial b_0}{\partial X}+h\Big(\frac{\partial^2}{\partial X^2}+\frac{\partial^2 }{\partial Z^2}\Big)a_0
				+ \frac{2gk}{\omega}\frac{\partial}{\partial X} |b_1|^2=0.
			\end{aligned}
		\end{equation}
		Let us replace $b_1\rightarrow \psi$, $a_0\rightarrow \varphi$, $b_0\rightarrow \rho$, $T\rightarrow t$, $X\rightarrow z$, $Z\rightarrow x$, which gives
		\begin{equation}\label{BR}
			\begin{aligned}
				&i\psi_t + \frac{\epsilon}{2}\omega_{22} \psi_{xx} + \frac{\epsilon}{2}\omega_{11} \psi_{zz} = \epsilon |\psi|^2\psi
				+\frac{\epsilon}{2g\omega}(g^2k^2-\omega^4)\rho\psi+\epsilon k \varphi_z\psi,   \\
				&\rho_t-c_g\rho_z +h( \varphi_{xx}+\varphi_{zz}) + \frac{2gk}{\omega} (|\psi|^2)_z=0,  \\
				&\varphi_t -c_g  \varphi_z +g\rho+ \frac{1}{\omega^2} (g^2 k^2-\omega^4)|\psi|^2=0.
			\end{aligned}
		\end{equation}
		Hence it is seen that  seen that the BR system identical to the Zakharov--Rubenchik (ZR) system also given in \cite{Ponce2005}
		\begin{equation}\label{ZR}
			\begin{aligned}
				&i\psi_t + \sigma_1 \psi_{xx} + \epsilon \psi_{zz} = \sigma_2 |\psi|^2\psi +W\rho\psi+WD\varphi_z\psi,   \\
				&\rho_t+\sigma_3\rho_z + \varphi_{xx}+\varphi_{zz} + D (|\psi|^2)_z=0,  \\
				&\varphi_t +\sigma_3 \varphi_z + \frac{1}{M}\rho + |\psi|^2=0.
			\end{aligned}
		\end{equation}
		We  refer to \cite{Ponce2005} for the physical meanings of the parameters and the dependent variables in \eqref{BRor} and \eqref{ZR}.
		
		The BR system \eqref{BRor} was derived in 1969 by Benney and Roskes in the context of gravity waves \cite{Benney1969}. Zakharov and Rubenchik \cite{zakharov1972nonlinear} derived the ZR system describing the interaction of a spectrally narrow small amplitude high frequency wave packet with a low frequency oscillations of acoustic type. We take the ZR system \eqref{ZR}  from \cite{Ponce2005}, and
		the derivation and physical background of the system can be found  in \cite{Zakharov1997} with details.

		Saut and Ponce \cite{Ponce2005} reduced the BR/ZR system to a nonlinear Schrodinger equation with nonlinear terms involving nonlocal terms and derivatives of the unknown and they studied  the local well-posedness of the Cauchy problem.
		Luong et al. handles  the Cauchy problem again for the two and three-dimensional Zakharov--Rubenchik system and they analyse its perturbation by a line soliton \cite{Luong2018}.
		Quintero and Cordero  worked the nonlinear orbital instability of ground state standing waves for a Benney--Roskes/Zakharov--Rubenchik system in two and three spatial directions \cite{Quintero2020}. In \cite{Martinez} Martinez and Palacios investigated the decay properties for the solutions to the initial value problem of the 1-D ZR/BR system.

		Ref. \cite{Luong2018} states that,  in suitable limits,  ZR (or BR) system contains the classical  Zakharov system  and the Davey-Stewartson systems \cite{Luong2018}. The Zakharov system is introduced in \cite{Zakharov1972} to describe the propagation of Langmuir waves in plasma.  Ref. \cite{CorderoCeballos2016} includes a rigorous justification of the Zakharov limit of the ZR system and \cite{oliveira2008adiabatic} presents the Schrödinger limit of the ZR system in one spatial dimension case for well-prepared initial data \cite{Luong2018}. Refs.   \cite{oliveira2008adiabatic} and \cite{Linares2009} are to be mentioned for the well-posedness results in the one dimensional setting.

		The Davey--Stewartson (DS) system was first derived by Davey and Stewartson in \cite{daveystewartson}. The DS system describing the propagation of two-dimensional water waves moving under the force of gravity in waters of finite depth is also a special case of the  Benney--Roskes system. Djordjevic and Redekopp extended the study of Davey and Stewartson to include the effects of capillarity \cite{Angeles1977}.
		In \cite{Ghidaglia1990}, Ghidaglia and Saut studied  the well-posedness of the Cauchy problem for DS equation. In the elliptic-hyperbolic and hyperbolic-hyperbolic  cases of the DS system, Linares showed in \cite{linares1993davey} that the initial value problem with small assumptions on the data is locally well-posed in weighted Sobolev space.

The symmetry algebra of the Davey--Stewartson system  is identified by Champagne and Winternitz in \cite{Champagne1988} as an inifinite-dimensional algebra of Kac--Moody--Virasoro (KMV) type, in the case when the system is integrable. Admitting a KMV algebra as the invariance algebra is a property which is seen in some other integrable equations in (2+1)-dimensions. Based on and motivated by the recent works concentrating on the BR/ZR system, we aim at studying the Lie symmetry algebra of the BR/ZR system to discover its group-theoretical properties and to make a comparison with the Lie symmetry algebra of the Davey--Stewartson system of equations. Besides, we search for some exact solutions. Briefly, we summarize the achievements of our work as follows.
		
		\begin{itemize}
			\item The symmetry algebra of the BR/ZR system is identified and compared with that of the Davey--Stewartson system.
			\item By a traveling wave ansatz, we obtain several exact solutions in terms of trigonometric, hyperbolic and elliptic functions.
			\item Besides many of the exact solutions being periodic, we obtain a line soliton solution the first time,
			to the best of our knowledge, that makes a good reference to the available literature.
            \item We also obtain a lump-type stationary solution.
		\end{itemize}
Section 2 is devoted to investigate the Lie symmetry algebra and in Section 3 we search for the exact solutions.
		
		\section{The Symmetry Algebra}
		
		\subsection{DS symmetry algebra}

        Before introducing our results on the BR/ZR system, we would like to put a remark on the Lie symmetry algebra of the Davey--Stewartson system. Following the work \cite{Champagne1988}, we consider the Davey--Stewartson system in the form
		\begin{equation}\label{DSW}
			\begin{aligned}
				 i\psi_t+\psi_{xx}+\epsilon_1\psi_{yy}&=\epsilon_2|\psi|^2\psi+\psi \phi, \\
				  \phi_{xx}+\delta_1 \phi_{yy}&=\delta_2(|\psi|^2)_{yy}.
			\end{aligned}
		\end{equation}
        This system was first derived in \cite{daveystewartson}, appearing as their equation (2.24a,b).
		 Considering $\psi=u+iv$, $\epsilon_1=\mp 1$, $\epsilon_2=\mp1$,
		the Lie symmetry algebra of the DS system \eqref{DSW} is reported in \cite{Champagne1988} to be generated by
		\begin{equation}
			 \bar{\mathcal{V}}=\bar{\mathcal{X}}(f)+\bar{\mathcal{Y}}(g)+\bar{\mathcal{Z}}(h)+\bar{\mathcal{W}}(m)
		\end{equation}
		with
		\begin{equation}\label{genW}
			\begin{aligned}
				\bar{\mathcal{X}}(f)&=f(t)\dt+\frac{\dot f(t)}{2} (x\dx+y\dy-u\du-v\dv-2\phi\df)     \\
				&-\frac{x^2+\epsilon_1 y^2}{8}\ddot{f}(t)(v\du-u\dv)
				-\frac{\dddot f(t)}{8}(x^2+\epsilon_1 y^2)\df,\\
				 \bar{\mathcal{Y}}(g)&=g(t)\dx-\frac{x}{2}\dot{g}(t)(v\du-u\dv)-\frac{x}{2}\ddot{g}(t)\df,\\
				\bar{\mathcal{Z}}(h)&=h(t)\dy-\frac{\epsilon_1 y}{2}\dot{h}(t)(v\du-u\dv)-\frac{\epsilon_1 y}{2}\ddot{h}(t)\df,\\
				\bar{\mathcal{W}}(m)  &=m(t) (v\du-u\dv)+\dot m(t)\df.
			\end{aligned}
		\end{equation}
		where $g(t)$, $h(t)$, $m(t)$ are arbitrary functions and $f(t)$ is stated to satisfy
		\begin{equation}
			f(t) = \begin{cases}
				\text{arbitrary,} &  \delta_1=-\epsilon_1=\mp1, \\
				a t^2+b t+c, &  \delta_1 \neq -\epsilon_1.
			\end{cases}
		\end{equation}
One of the main results of \cite{Champagne1988} is that, exactly in the integrable case $\delta_1=-\epsilon_1$ of the DS system,  the  symmetry algebra of the DS system \eqref{DSW} is an infinite-dimensional Lie algebra of
Kac--Moody--Virasoro type, generated by vector fields in \eqref{genW} depending on four arbitrary functions of time. The algebra has the Levi decomposition $\bar{\mathcal{X}}(f) \semi \{\bar{\mathcal{Y}}(g),\bar{\mathcal{Z}}(h),\bar{\mathcal{W}}(m) \}$ \cite{Champagne1988}. For the structural information on KMV algebras and further examples of PDEs enjoying the KMV algebra as the invariance algebra we can refer to \cite{gungor2006virasoro}.
		
		Before deriving the system \eqref{DSW}, the first set of equations Davey and Stewartson derive in their work \cite{daveystewartson}, which are the equations (2.14)-(2.15) of \cite{daveystewartson}, is a system of the form
		\begin{subequations}\label{DS}
			\begin{eqnarray}
				\label{DSa}  i\psi_t+\psi_{xx}+\epsilon_1\psi_{yy}&=&\epsilon_2|\psi|^2\psi+\psi \varphi_y, \\
				\label{DSb}  \varphi_{xx}+\delta_1\varphi_{yy}&=&\delta_2(|\psi|^2)_{y}
			\end{eqnarray}
		\end{subequations}
		where $\delta_1,\delta_2,\epsilon_1$ and $\epsilon_2$ are real constants. Transition from \eqref{DS} to \eqref{DSW} is straightforward by differentiating \eqref{DSb} with respect to $y$ and replacing $\varphi_y=\phi$ in both of the equations. Denoting $\psi=u+iv$ and assuming  $\epsilon_1=\mp 1$, $\epsilon_2=\mp1$, we determined that the Lie symmetry algebra of \eqref{DS}  is generated by the vector field
		\begin{equation}
			 \mathcal{V}=\mathcal{X}(\tau)+\mathcal{Y}(\alpha)+\mathcal{Z}(\beta)+\mathcal{W}(\mu)+\mathcal{R}(\gamma)+\mathcal{S}(\delta)
		\end{equation}
		with
		\begin{equation}\label{gen}
			\begin{aligned}
				\mathcal{X}(\tau)&=\tau(t)\dt+\frac{\dot \tau(t)}{2} (x\dx+y\dy-u\du-v\dv-\varphi\dvf)    \\
				&-\frac{x^2+ \epsilon_1 y^2}{8}\ddot{\tau}(t)(v\du-u\dv)
				-\frac{\dddot\tau(t)}{8}(x^2y+\frac{\epsilon_1}{3} y^3)\dvf,\\
				 \mathcal{Y}(\alpha)&=\alpha(t)\dx-\frac{x}{2}\dot{\alpha}(t)(v\du-u\dv)-\frac{xy}{2}\ddot{\alpha}(t)\dvf,\\
				\mathcal{Z}(\beta)&=\beta(t)\dy-\frac{ y}{2\epsilon_1}\dot{\beta}(t)(v\du-u\dv)+\frac{\ddot{\beta}(t)}{4 \epsilon_1}(\delta_1x^2- y^2)\dvf,\\
				\mathcal{W}(\mu)  &=\mu(t) (v\du-u\dv)+y\dot\mu(t)\dvf, \\
				\mathcal{R}(\gamma)& = x \gamma(t) \dvf,\\
				\mathcal{S}(\delta)&= \delta(t)\dvf.
			\end{aligned}
		\end{equation}
		where $\alpha(t)$, $\beta(t)$, $\mu(t)$, $\gamma(t)$, $\delta(t)$ are arbitrary functions whereas the constants $\epsilon_1$, $\delta_1$ and the function $\tau(t)$ must satisfy
		\begin{equation}
			(\epsilon_1+\delta_1)\dddot{\tau}(t)=0.
		\end{equation}
		Therefore we have
		\begin{equation}
			\tau(t) = \begin{cases}
				\text{arbitrary,} &  \delta_1=-\epsilon_1=\mp1, \\
				\tau_2 t^2+\tau_1 t+\tau_0, &  \delta_1 \neq -\epsilon_1.
			\end{cases}
		\end{equation}		
		For the generators \eqref{gen}, the nonzero commutations are
		\begin{equation}
			\begin{aligned}\label{kom1}
				&[\mathcal{X}(\tau_1),\mathcal{X}(\tau_2)]=\mathcal{X}(\tau_1\dot\tau_2 -\dot\tau_1 \tau_2),  &&[\mathcal{X}(\tau),\mathcal{Y}(\alpha)]=\mathcal{Y}(\tau \dot\alpha-\frac{1}{2}\dot\tau \alpha ), \\
                &[\mathcal{X}(\tau),\mathcal{Z}(\beta)]=\mathcal{Z}(\tau \dot\beta-\frac{1}{2}\dot\tau \beta ),   	
				&&[\mathcal{X}(\tau),\mathcal{W}(\mu)]=\mathcal{W}(\tau \dot\mu ), \\
				&[\mathcal{X}(\tau),\mathcal{R}(\gamma)]=\mathcal{R}(\tau\dot\gamma+\dot\tau\gamma ),
				&&[\mathcal{X}(\tau),\mathcal{S}(\delta)]=\mathcal{S}(\tau \dot\delta+\frac{1}{2}\dot\tau \delta ), \\
                &[\mathcal{Y}(\alpha_1),\mathcal{Y}(\alpha_2)]=-\frac{1}{2}\mathcal{W}(\alpha_1 \dot\alpha_2-\dot\alpha_1\alpha_2 ),
				 &&[\mathcal{Z}(\beta_1),\mathcal{Z}(\beta_2)]=-\frac{\epsilon_1}{2}\mathcal{W}(\beta_1 \dot\beta_2-\dot\beta_1\beta_2  ), 	\\
				&[\mathcal{Y}(\alpha),\mathcal{Z}(\beta)]=\frac{1}{2}\mathcal{R}( \ddot\alpha \beta+\delta_1\epsilon_1\alpha \ddot\beta ),
				&&[\mathcal{Y}(\alpha),\mathcal{R}(\gamma)]=\mathcal{S}(\alpha \gamma ), \\  		
				&[\mathcal{Z}(\beta),\mathcal{W}(\mu)]=\mathcal{S}(\beta \dot\mu ).						
			\end{aligned}
		\end{equation}
		We can summarize this finding as follows.			
		\begin{prop}
		When $\delta_1=-\epsilon_1=\mp1$, the symmetry algebra of the DS system \eqref{DS} is an infinite-dimensional Kac--Moody--Virasoro algebra generated by the vector fields given in \eqref{gen}, including 6 arbitrary functions of time. The algebra has the Levi decomposition
\begin{equation}
\{\mathcal{X}(\tau)\}\semi \{\mathcal{Y}(\alpha),\mathcal{Z}(\beta),\mathcal{W}(\mu),\mathcal{R}(\gamma),\mathcal{S}(\delta)\}.
\end{equation}
        \end{prop}
\begin{rmk}
The Lie algebra of the DS system in \eqref{DSW} includes 4 arbitrary functions of time if $\delta_1=-\epsilon_1$, and 3 arbitrary functions of time if $\delta_1\neq -\epsilon_1$. The Lie algebra of the DS system in \eqref{DS} includes 6 arbitrary functions of time if $\delta_1=-\epsilon_1$, and 5 arbitrary functions of time if $\delta_1\neq -\epsilon_1$; hence richer than that of \eqref{DSW}.
\end{rmk}

		\subsection{BR/ZR symmetry algebra}
	    From \eqref{BR} we solve
		\begin{equation}
			\rho = \frac{1}{g} \Big[c_g \varphi_z-\varphi_t - \frac{1}{\omega^2}(g^2k^2-\omega^4) |\psi|^2    \Big]
		\end{equation}
		and get
        {\small
        \begin{equation}
		\begin{aligned}
			i\psi_t &+ \frac{\epsilon}{2}\omega_{22} \psi_{xx} + \frac{\epsilon}{2}\omega_{11} \psi_{zz} = \epsilon(1-\frac{A_0^2}{2g^2 \omega^3}) |\psi|^2\psi
			-\frac{\epsilon}{2g^2 \omega}\psi\Big\{A_0\frac{\partial}{\partial t}-(A_0 c_g+2g^2k\omega)\frac{\partial}{\partial z}\Big\}\varphi,  \\
			 \varphi_{xx}&+(1-\frac{c_g^2}{gh})\varphi_{zz}-\frac{1}{gh}\varphi_{tt}+\frac{2c_g}{gh}\varphi_{zt}=
			\frac{1}{gh\omega^2}\Big\{A_0 \frac{\partial}{\partial t}-(A_0 c_g+2g^2k\omega)\frac{\partial}{\partial z}\Big\}|\psi|^2
		\end{aligned}
        \end{equation}}
		with $A_0=g^2k^2-\omega^4$.
		Similarly, we solve from \eqref{ZR} to obtain
		\begin{equation}
			\rho = -M(\varphi_t+\sigma_3\varphi_z+|\psi|^2)
		\end{equation}
		and find
		\begin{equation}
			\begin{split}
				&i\psi_t+\sigma_1 \psi_{xx}+\epsilon\psi_{zz}=(\sigma_2-WM)|\psi|^2\psi-W \psi\Big\{M\frac{\partial}{\partial t}+(M\sigma_3-D)\frac{\partial}{\partial z}\Big\}\varphi,\\
				&\varphi_{xx}+(1-M\sigma_3^2)\varphi_{zz}-M\varphi_{tt}-2 M \sigma_3 \varphi_{zt}=\Big\{M\frac{\partial}{\partial t}+(M\sigma_3-D)\frac{\partial}{\partial z}\Big\} |\psi|^2.
			\end{split}
		\end{equation}
See the correspondence between the new form of the BR and ZR systems. We shall concentrate on this final form of the ZR system in calculating the symmetries. Let us replace $z\rightarrow y$ and re-label the constants as follows
\begin{align}
				\label{BRZRa} &i\psi_t+a\psi_{xx}+\epsilon_0\psi_{yy}=(b-wm)|\psi|^2\psi-w \psi\Big\{m\frac{\partial}{\partial t}+(mc-d)\frac{\partial}{\partial y}\Big\}\varphi, \\
				\label{BRZRb} &\varphi_{xx}+(1-mc^2)\varphi_{yy}-m\varphi_{tt}-2 m c \varphi_{yt}=\Big\{m\frac{\partial}{\partial t}+(mc-d)\frac{\partial}{\partial y}\Big\} |\psi|^2.
\end{align}		
Clearly, when $m=0$, this system reduces to \eqref{DS}. We would like to point out here that $a,b,c$ are considered  as nonzero real numbers and $\epsilon_0,d,m,w$  as positive.

We  evaluate the symmetries in two cases. First, we shall find the symmetry algebra for the system \eqref{BRZRa}-\eqref{BRZRb}, so as to make a comparison with the algebra of Eq.  \eqref{DS}. Afterwards, we are going to simplify the derivatives appearing on the right hand sides, and that will give the chance to see also the comparison with the symmetry algebra of Eq. \eqref{DSW}.
		\subsubsection{BR/ZR  in the first form}
		Separating  $\psi=u+iv$,
		the Lie symmetry algebra  BR/ZR system \eqref{BRZRa}-\eqref{BRZRb}
		is generated by the infinitesimal
		\begin{equation}
			\mathcal{V}=\tau_0\dt +\xi_0 \dx+\eta_0 \dy -\chi(t)(v\du-u\dv)+\Big[\frac{1}{mw}\chi(t)+P(x,\zeta)\Big]\dvf
		\end{equation}
		where $\tau_0$, $\xi_0$, $\eta_0$ and
		\begin{equation}
			\zeta=my-(m c-d)t,
		\end{equation}
		with $\chi(t)$ and $P(x,\zeta)$ satisfying
		\begin{equation}
			P_{xx}+m (m-d^2)P_{\zeta \zeta}=\frac{1}{w}\ddot{\chi}(t).
		\end{equation}
		The RHS of this equation depends on $t$ only, and the LHS depends on $x$ and $\zeta$. Therefore, it must be equal to a constant, say, $c_1$,
		\begin{equation}\label{eqP}
			P_{xx}+m (m-d^2)P_{\zeta \zeta}=\frac{1}{w}\ddot{\chi}(t)=c_1.
		\end{equation}
		Therefore we have $\displaystyle \chi(t)=\frac{c_1w}{2}t^2+s_1 t+s_0$. For the solution to $P_{xx}+m (m-d^2)P_{\zeta \zeta}=c_1$, let $\displaystyle P(x,\zeta)=\bar{P}(x,\zeta)+\frac{c_1}{2} x^2$. Then $\bar{P}$ satisfies
		\begin{equation}\label{eqP}
			\bar{P}_{xx}+m (m-d^2)\bar{P}_{\zeta \zeta}=0.
		\end{equation}
		
		\noindent\textbf{(A)} Suppose $m<d^2$. Set $\displaystyle\alpha_0^2=m (d^2-m)$. Then
		\begin{equation}
			\bar{P}_{xx}-\alpha_0^2\bar{P}_{\zeta \zeta}=0
		\end{equation}
		is integrated to
		\begin{equation}
			\bar{P}=F(\alpha_0 x-\zeta) + G(\alpha_0 x+\zeta)
		\end{equation}
		and hence
		\begin{equation}
			P(x,\zeta)=F(\alpha_0 x-\zeta) + G(\alpha_0 x+\zeta)+\frac{c_1}{2} x^2.
		\end{equation}
		We see that the Lie algebra is spanned by
		\begin{equation}
			\begin{aligned}
				\mathcal{X}_1&=\dt, \quad \mathcal{X}_2 = \dx, \quad  \mathcal{X}_3 = \dy, \\
				\mathcal{X}_4&=-v\du +u \dv+\frac{1}{mw}\dvf, \\
				\mathcal{X}_5&=t(-v\du +u \dv)+\frac{t}{mw}\dvf,\\
				\mathcal{X}_6&=t^2(-v\du +u \dv)+(\frac{t^2}{mw}+\frac{x^2}{w})\dvf,\\
				\mathcal{P_A}(F)&=F\Big(\sqrt{m(d^2-m)}\,x-my+(mc-d)t\Big)\dvf,\\
				\mathcal{Q_A}(G)&=G\Big(\sqrt{m(d^2-m)}\,x+my-(m c-d)t\Big)\dvf
			\end{aligned}
		\end{equation}
		where $F(s)$ and $G(s)$ are arbitrary single-variable functions. The nonzero commutations are		
		\begin{equation}
			\begin{aligned}\label{kom2}
				&[\mathcal{X}_1,\mathcal{X}_5]= \mathcal{X}_4,     &&[\mathcal{X}_1,\mathcal{X}_6]=2 \mathcal{X}_5, \\
				&[\mathcal{X}_1,\mathcal{P_A}(F)]=(mc-d)\mathcal{P_A}(F'),    &&[\mathcal{X}_1,\mathcal{Q_A}(G)]=(d-mc)\mathcal{Q_A}(G'),  \\
				&[\mathcal{X}_2,\mathcal{P_A}(F)]=\sqrt{m(d^2 -m)}\,\mathcal{P_A}( F'),    &&[\mathcal{X}_2,\mathcal{Q_A}(G)]=\sqrt{m(d^2 -m)}\,\mathcal{Q_A}( G'),  \\
				&[\mathcal{X}_3,\mathcal{P_A(F)}]=-m\mathcal{P_A}( F'),  &&[\mathcal{X}_3,\mathcal{Q_A}(G)]=m\mathcal{P_A}( G'),\\
				&[\mathcal{X}_2,\mathcal{X}_6]= \frac{1}{w \sqrt{m(d^2 -m)}} (\mathcal{P_A}|_{F(s)=s} +\mathcal{Q_A}|_{G(s)=s}).
			\end{aligned}
		\end{equation}
		
		\noindent\textbf{(B)}
		Suppose $m>d^2$. $\bar{P}(x,\zeta)$ satisfies
		\begin{equation}
			\bar{P}_{xx}+m (m-d^2)\bar{P}_{\zeta \zeta}=0.
		\end{equation}
		Let us scale $\displaystyle \hat \zeta = \frac{\zeta}{\sqrt{m (m-d^2)}}$. Then we have
		\begin{equation}
			\bar{P}_{xx}+\bar{P}_{\hat{\zeta} \hat{\zeta}}=0.
		\end{equation}
		Then
		\begin{equation}
			\bar{P}=H(x,\hat{\zeta}),
		\end{equation}
		where $H$ is any harmonic function. Therefore we find that
		\begin{equation}
			P(x,\zeta)=H\Big(x,\frac{\zeta}{\sqrt{m (m-d^2)}}\Big)+\frac{c_1}{2} x^2.
		\end{equation}
		Similarly,  we obtain the previous generators $\mathcal{X}_i$, $i=1,...,6$ and
		\begin{equation}
			\mathcal{P_B}(H)=H(x,\sigma)\dvf,  \qquad \sigma=\frac{my-(m c-d)t}{\sqrt{m(m-d^2)}}\\
		\end{equation}
		where $H$ is an arbitrary harmonic function. The nonzero commutation relations are		
		\begin{equation}
			\begin{aligned}\label{kom2ex2}
                &[\mathcal{X}_1,\mathcal{X}_5]= \mathcal{X}_4,     &&[\mathcal{X}_1,\mathcal{X}_6]=2 \mathcal{X}_5, \\
                &[\mathcal{X}_2,\mathcal{X}_6]=\mathcal{P_B}(H)|_{H(x,\sigma)=2x/w},
				&&[\mathcal{X}_1,\mathcal{P_B}(H)]= \frac{d-mc}{\sqrt{m(m-d^2)}}\, \mathcal{P_B}\Big(  H_{\sigma}(x,\sigma)\Big), \\ &[\mathcal{X}_2,\mathcal{P_B}(H)]=\mathcal{P_B}\Big(H_{x}(x,\sigma)\Big),
				&&[\mathcal{X}_3,\mathcal{P_B}(H)]=\sqrt{\frac{m}{m-d^2}} \, \mathcal{P_B}\Big(H_\sigma (x,\sigma)\Big).
			\end{aligned}
		\end{equation}

		\noindent\textbf{(C)}
		Suppose $m=d^2$. Then solving $P_{xx}(x,\zeta)=c_1$ yields
		\begin{equation}
			P(x,\zeta)=\frac{c_1}{2} x^2+R(\zeta) x+S(\zeta),
		\end{equation}
		where $R$, $S$ are arbitrary functions. In addition to $\mathcal{X}_i$, $i=1,...,6$, we obtain the generators
		\begin{equation}
        \begin{aligned}
			&\mathcal{P_C}(R)=xR\big(dy-(cd-1)t\big)\dvf,\\
			&\mathcal{Q_C}(S)=S\big(dy-(cd-1)t\big)\dvf
		\end{aligned}
        \end{equation}
		where $R(s)$ and $S(s)$ are arbitrary functions of a single variable. We present the nonzero commutation relations as follows.
		\begin{equation}	
			\begin{aligned}\label{kom2ex}
                &[\mathcal{X}_1,\mathcal{X}_5]= \mathcal{X}_4,     &&[\mathcal{X}_1,\mathcal{X}_6]=2 \mathcal{X}_5, \\
				&[\mathcal{X}_1,\mathcal{P_C}(R)]=(1-cd)\mathcal{P_C}( R') ,   &&[\mathcal{X}_2,\mathcal{P_C}(R)]= \mathcal{Q_C}(R), \\
				&[\mathcal{X}_3,\mathcal{P_C}(R)]=d \mathcal{P_C}(R'),    &&[\mathcal{X}_1,\mathcal{Q_C}(S)]=(1-cd)\mathcal{Q_C}(S')  \\
				&[\mathcal{X}_3,\mathcal{Q_C}(S)]= d \mathcal{Q_C}(S'), &&[\mathcal{X}_2,\mathcal{X}_6]=\mathcal{P_C}(R)|_{R(s)=2/w}.\\
			\end{aligned}
		\end{equation}
		\begin{prop}
        The invariance algebra of the BR/ZR system \eqref{BRZRa}-\eqref{BRZRb} is infinite-dimensional and has the following bases.
        \begin{itemize}
        \item[$\circ$] $m<d^2: \qquad\mathcal{L_A}=\{\mathcal{X}_i,\mathcal{P_A}(F),\mathcal{Q_A}(G)\}, \quad i=1,\ldots, 6,$
        \item[$\circ$] $m>d^2: \qquad\mathcal{L_B}=\{\mathcal{X}_i,\mathcal{P_B}(H)\}, \qquad \qquad i=1,\ldots, 6,$
        \item[$\circ$] $m=d^2: \qquad\mathcal{L_C}=\{\mathcal{X}_i,\mathcal{P_C}(R),\mathcal{Q_C}(S)\}, \quad i=1,\ldots, 6.$
        \end{itemize}
        \end{prop}
		
		\subsubsection{BR/ZR in the second form}
		Notice that the directional derivatives on the RHS of Eqs. \eqref{BRZRa}-\eqref{BRZRb} are in the same direction. This gives us the chance to simplify the right hand sides of these equations by a rotational transformation. We keep the variables $(x,\psi,\varphi)$ as they are and do the transformation $(y,t)\rightarrow (\xi,\eta)$ with
		\begin{equation}
			t=k_1 \xi+k_2\eta, \qquad y=\ell_1\xi+\ell_2 \eta.
		\end{equation}
		with the condition $\kappa_0=k_1\ell_2-\ell_1k_2\neq 0$. When we choose $k_1=m$, $\ell_1=mc-d$, the mentioned directional derivative becomes
		\begin{equation}
			m\frac{\partial}{\partial t}+(mc-d)\frac{\partial}{\partial y}=\frac{\partial}{\partial \xi}.
		\end{equation}
		With this transformation, we write \eqref{BRZRa}  as
		\begin{equation}\label{eq1}
			i\big(\frac{-\ell_1}{\kappa_0}\big)\psi_{\eta} +a \psi_{xx}  + \frac{\epsilon_0 k_2^2}{\kappa_0^2} \psi_{\xi\xi} + \frac{\epsilon_0 k_1^2}{\kappa_0^2} \psi_{\eta\eta}
			-\frac{2\epsilon_0 k_1 k_2}{\kappa_0^2}\psi_{\xi\eta}+i\big(\frac{\ell_2}{\kappa_0}\big)\psi_{\xi}=(b-wm) |\psi^2|\psi-w  \psi \varphi_\xi.
		\end{equation}
		This way, the term $\varphi_{\xi\eta}$ appearing on the LHS of  \eqref{BRZRb} is eliminated if we choose $k_2=d\sqrt{m}$, $\ell_2=\sqrt{m}(c d-1)$ and we obtain
		\begin{equation}\label{eq2}
			\varphi_{xx}+\frac{1}{d^2-m} (\varphi_{\xi\xi}-\varphi_{\eta\eta})=(|\psi|^2)_{\xi}
		\end{equation}
		with the condition $\kappa=\sqrt{m}(d^2-m)\neq 0$. Differentiating \eqref{eq2} with respect to $\xi$ once and replacing $\varphi_\xi=\phi$, we consider \eqref{eq1} and \eqref{eq2} in the form
		\begin{align}
			\label{BRZR2a}    ia_1\psi_{t} +a \psi_{xx}  + a_2 \psi_{yy} + a_3 \psi_{tt}
			+a_4\psi_{ty}+ia_5\psi_{y}&=(b-wm) |\psi^2|\psi-w   \psi \phi,\\
			\label{BRZR2b}   \phi_{xx}+\frac{1}{d^2-m} (\phi_{yy}-\phi_{tt})&=(|\psi|^2)_{yy}
		\end{align}
		with
		\begin{equation}
			\begin{aligned}
				&a_1=\frac{d-mc}{(d^2-m)\sqrt{m}}, 	 &&a_2=\frac{\epsilon_0 d^2}{(d^2-m)^2}, \quad a_3=\frac{\epsilon_0 m}{(d^2-m)^2},  \\ &a_4=-\frac{2\epsilon_0 d \sqrt{m}}{(d^2-m)^2},      &&a_5=\frac{dc-1}{d^2-m}.
			\end{aligned}
		\end{equation}
		The symmetries are generated by the vector fields
		\begin{align}
			&\bar{\mathcal{X}}_1=\dt, \quad \bar{\mathcal{X}}_2 = \dx, \quad  \bar{\mathcal{X}}_3 = \dy, \quad \bar{\mathcal{X}}_4=-v\du +u \dv, \\
			&\bar{\mathcal{X}}_5=(dt+\sqrt{m}y)(-v\du +u \dv)+\frac{1}{w\sqrt{m}}\df,\\
			&\bar{\mathcal{X}}_6=(dt+\sqrt{m}y)^2(-v\du +u \dv)+\frac{2(dt+\sqrt{m}y) }{w\sqrt{m}}\df
		\end{align}
		with the nonzero commutations being		
		\begin{equation}
			\begin{aligned}\label{kom3}
				&[\bar{\mathcal{X}}_1,\bar{\mathcal{X}}_5]=d \bar{\mathcal{X}}_4,   &&[\bar{\mathcal{X}}_1,\bar{\mathcal{X}}_6]=2 d \bar{\mathcal{X}}_5, \\
				&[\bar{\mathcal{X}}_3,\bar{\mathcal{X}}_5]=\sqrt{m} \bar{\mathcal{X}}_4, &&[\bar{\mathcal{X}}_3,\bar{\mathcal{X}}_6]=2 \sqrt{m} \bar{\mathcal{X}}_5.
			\end{aligned}
		\end{equation}
\begin{prop}
The BR/ZR system \eqref{BRZR2a}-\eqref{BRZR2b} admits a six-dimensional Lie algebra with the basis $\{\bar{\mathcal{X}}_i\}$, $i=1,\ldots,6$.
\end{prop}	
		
		\section{Exact Solutions}
		A  solution of the form
\begin{equation}\label{ansatz}
			\psi =e^{i\theta}\Psi(\nu),  \qquad \varphi= \Phi (\nu), \quad \nu=\nu_1 x+\nu_2y+\nu_0t, \quad \theta=\theta_1 x+\theta_2 y+\theta_0 t,
		\end{equation}
where $\nu_i$, $\theta_i$, $i=0,1,2$  are real constants \cite{Elvis}, reduces  the  BR/ZR system  in \eqref{BRZRa}-\eqref{BRZRb} to the system of nonlinear ordinary differential equations
		\begin{align}
			(a\nu_1^2+\epsilon_0  \nu_2^2 )\Psi'' &+i(2a\nu_1 \theta_1+ 2 \epsilon_0 \nu_2 \theta_2+\nu_0)\Psi' -( a \theta_1^2+\epsilon_0 \theta_2^2 +\theta_0)\Psi  \nonumber \\
			&+( m w-b)\Psi^3+w( m\nu_0+mc\nu_2 -d\nu_2) \Psi \Phi'  =0,   \label{eqpsi}\\
			2(d\nu_2-m\nu_0-mc\nu_2 )\Psi \Psi'  &+\big( \nu_1^2+\nu_2^2-m(\nu_0+c\nu_2)^2 \big)\Phi''=0
		\end{align}
which requires
		\begin{align}
			2a\nu_1\theta_1+ 2 \epsilon_0 \nu_2 \theta_2+\nu_0=0.
		\end{align}
		Integration of second equation one time and taking the constant of integration to be zero, we obtain
		\begin{equation}\label{eqfi}
			\Phi'=K \Psi^2, \qquad K=\frac{m\nu_0 +(mc-d)\nu_2}{\nu_1^2+\nu_2^2-m(\nu_0+c\nu_2)^2}.
		\end{equation}
		Substituting $	\Phi'$ into first equation we obtain
		\begin{equation}\label{power22}
			\Psi''=2A \Psi^3 +B \Psi
		\end{equation}
		where
		\begin{equation}
			A=\frac{b-mw- wK(m\nu_0+(mc -d)\nu_2) }{2(a \nu_1^2+\epsilon_0 \nu_2^2)}, 	
			\qquad 	B= \frac{a \theta_1^2+\epsilon_0 \theta_2^2+\theta_0 }{a \nu_1^2+\epsilon_0 \nu_2^2 } \, .
		\end{equation}
		Multiplying both sides of \eqref{power22} by $\Psi' $, we obtain
		\begin{equation}\label{power}
			(\Psi')^2=A \Psi ^4+B\Psi^2+C
		\end{equation}
		where $C$ is an arbitrary constant.

Having obtained \eqref{power}, we are going to produce several types of solutions. First of all, by a scaling  $\Psi=\alpha_1 p$, $\nu=\beta_1 q$,  \eqref{power} can be converted to a specific canonical form that yields elliptic functions. We first assume $\Delta=B^2-4AC>0$ and present three different cases.

\vspace{.5cm}
\noindent \textbf{(i)} \   When  $A>0$, $B<0$, $C>0$, for $\alpha_1^2=C\beta_1^2$ and $\displaystyle \beta_1^2=-\frac{B+\sqrt{B^2-4AC}}{2AC}$, \eqref{power} is transformed to
\begin{equation}
\left(\frac{dp}{dq}\right)^2=(1-p^2)(1-\kappa^2p^2)
\end{equation}
with $\kappa^2=AC\beta_1^4$. We obtain
\begin{equation}\label{sn}
\Psi=\alpha_1 \mathrm{sn}\Big(\frac{1}{\beta_1}(\nu+\delta_0),\kappa\Big).
\end{equation}
To illustrate this elliptic periodic  solution at $t=0$, we choose all the parameters in \eqref{BRZRa}-\eqref{BRZRb} and \eqref{ansatz} equal to $1$, except  $\theta_0=-5$, $\nu_0=-4$, $C=0.5$. Thus  Figure 1 is the plot of

\begin{equation}
			\label{snplot}\Psi=0.625953\times \mathrm{sn}\Big(1.12965 (x + y), 0.418871\Big).
\end{equation}
\begin{figure*}[hbt!]
			\centering
			%\subfigure[]{\label{main:a}
             \includegraphics[scale=.55]{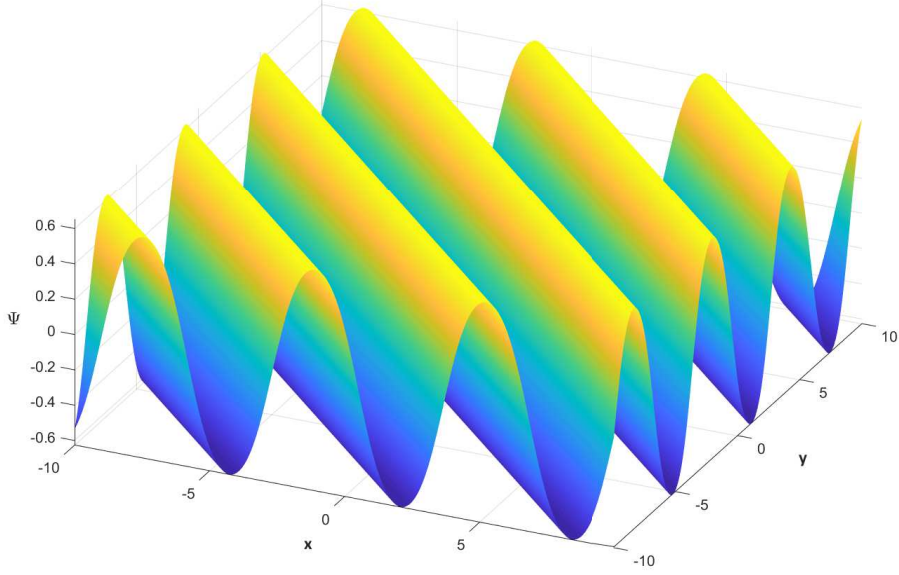}
			\caption{The plot for $\Psi$ of \eqref{sn} for $t=0$. The values of parameters are equal to $1$ except $\theta_0=-5$, $\nu_0=-4$, $C=0.5$.}
			\label{fig:main}
\end{figure*}

\noindent \textbf{(ii)} \   In case  $A<0$, $C>0$, with $\displaystyle \alpha_1^2=-\frac{B+\sqrt{B^2-4AC}}{2A}$, $\displaystyle \kappa^2=\frac{A\alpha_1^2}{2A\alpha_1^2+B}$ and  $\displaystyle\beta_1^2=-\frac{\kappa^2}{A\alpha_1^2}$, \eqref{power} is transformed to
\begin{equation}
\left(\frac{dp}{dq}\right)^2=(1-p^2)(\kappa'^2+\kappa^2p^2)
\end{equation}
with $\kappa'^2=1-\kappa^2$. Then we get
\begin{equation}
\Psi=\alpha_1 \mathrm{cn}\Big(\frac{1}{\beta_1}(\nu+\delta_0),\kappa\Big).
\end{equation}

\noindent \textbf{(iii)} \   When $A,B,C>0$, for $\alpha_1^2=C\beta_1^2$ and $\displaystyle \beta_1^2=\frac{B-\sqrt{B^2-4AC}}{2AC}$, \eqref{power} is transformed to
\begin{equation}
\left(\frac{dp}{dq}\right)^2=(1+p^2)(1+\kappa'^2p^2)
\end{equation}
with $\kappa'^2=AC\beta_1^4$. We obtain
\begin{equation}
\Psi=\alpha_1 \mathrm{tn}\Big(\frac{1}{\beta_1}(\nu+\delta_0),\kappa\Big).
\end{equation}
There are several possible types of elliptic solutions that can be obtained from  \eqref{power} through a similar treatment yet not to be included here. Furthermore, one can also obtain trigonometric and hyperbolic type solutions, which can also be recovered as the limiting cases of the elliptic solutions family.  In special, when $C=0$ and $B\neq 0$ in \eqref{power},  the equation reduces to
		\begin{equation}
			\frac{d \Psi}{\Psi \sqrt{A \Psi^2+B}} = \varepsilon d \nu
		\end{equation}
		where $\varepsilon =\mp 1$. We can evaluate this integral in the following three different cases.

\vspace{.5cm}		
\noindent \textbf{(iv)} \ In case $A>0$ and $B>0$,we obtain
		\begin{align}
			\Psi(\nu)&=-\varepsilon\sqrt{\frac{B}{A}}\ \mathrm{csch}\Big(\sqrt{B}(\nu+\delta_0)\Big),\\
			\Phi(\nu)&=-\frac{K\sqrt{B}}{A}\coth \Big(\sqrt{B}(\nu+\delta_0)\Big)+\Phi_0.
		\end{align}
		\textbf{(v)} \ When $A>0$ and $B<0$, we obtain
		\begin{align}
			\Psi(\nu)&=\sqrt{-\frac{B}{A}}\ \mathrm{sec}\Big(\sqrt{-B}(\nu+\delta_0)\Big), \\
			\Phi(\nu)&=\frac{K\sqrt{-B}}{A}\tan \Big(\sqrt{-B}(\nu+\delta_0)\Big)+\Phi_0.
		\end{align}
		\textbf{(vi)} \ If $A<0$ and $B>0$, we obtain
		\begin{align}
			\label{sech}\Psi(\nu) &=\sqrt{-\frac{B}{A}}\ \mathrm{sech}\Big(\sqrt{B}(\nu+\delta_0)\Big), \\
			\label{tanh}\Phi(\nu) &=-\frac{K\sqrt{B}}{A}\tanh \Big(\sqrt{B}(\nu+\delta_0)\Big)+\Phi_0.
		\end{align}
\begin{rmk}
For the (1+1)-dimensional case, \cite{Luong2018} and \cite{luong18thesis} present an exact solution of a similar form to (vi) and  in \cite{luong18thesis}, there are also numerical investigations  based on that solution. To the best of our knowledge, the solution \eqref{sech}-\eqref{tanh}, plotted in Figure 2, appear the first time in the literature for the (2+1)-dimensional BR/ZR system. We think this solution might also be useful for the kind of numerical investigations done for a stability analysis appearing in Chapter 6 of \cite{luong18thesis}.
\end{rmk}
To illustrate the solution in \eqref{sech}-\eqref{tanh}, we choose all the parameters in \eqref{BRZRa}-\eqref{BRZRb} and \eqref{ansatz} equal to $1$, except $w=3$ and $\theta_2=-3/2$. Therefore in Figure 2 we plot
       \begin{equation}
			\Psi =\sqrt{17}\ \mathrm{sech}\Big(\sqrt{\frac{17}{8}}(x+y)\Big), \quad \Phi =-\sqrt{34}\tanh \Big(\sqrt{\frac{17}{8}}(x+y)\Big).
		\end{equation}
\begin{figure*}[h!]		
			\centering
			\subfigure[]{\label{main:a}\includegraphics[scale=.45]{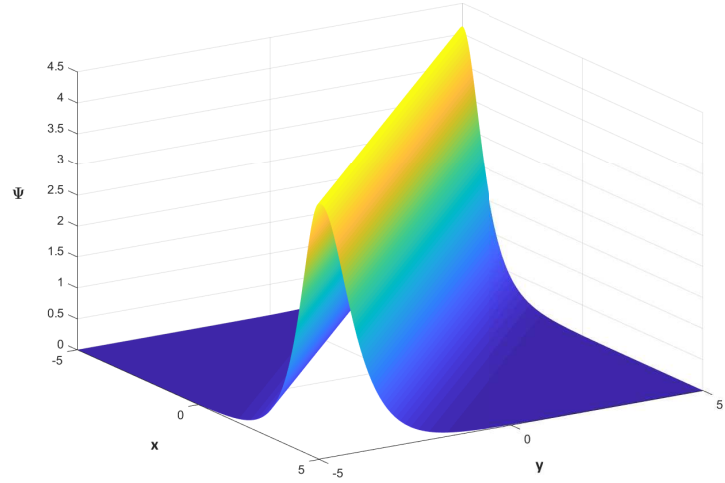}}
					\centering
			\subfigure[]{\label{main:b}\includegraphics[scale=.52]{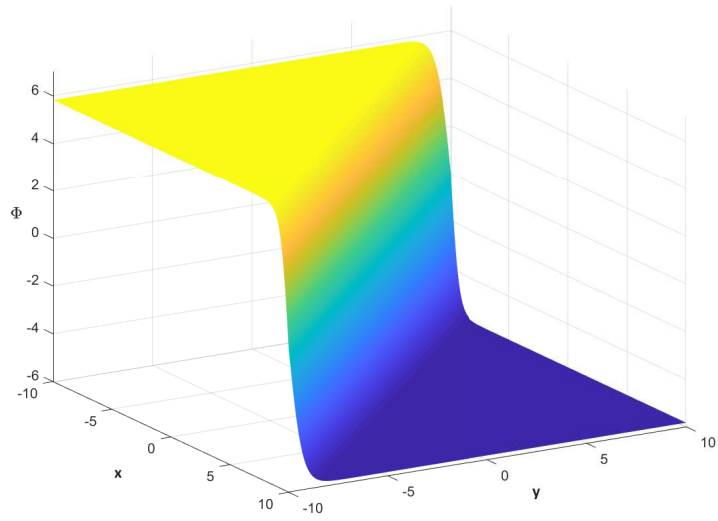}}
			\caption{Plots for (a) \eqref{sech}, (b) \eqref{tanh} for $t=0$. The values of parameters are equal to 1 except $w=3$ and $\theta_2=-3/2$.}
			\label{fig:main}
		\end{figure*}

If $ \Delta=B^2-4 A C=0$, then \eqref{power} takes the form
		\begin{equation}
			\frac{d\Psi}{2A\Psi^2+B}=\frac{\varepsilon}{2\sqrt{A}}\ d\nu
		\end{equation}
		where $\varepsilon =\mp 1$.

\vspace{.5cm}
\noindent \textbf{(vii)} \  If $B>0$, we find
		\begin{equation}
			\Psi(\nu)=\varepsilon\sqrt{\frac{B}{2A}}\tan\big( \sqrt{\frac{B}{2}}(\nu+\delta_0)\big)
		\end{equation}
		where $A$ and $C$ are positive, $\delta_0$ is integration constant. We integrate \eqref{eqfi} and find
		\begin{align}
			\Phi(\nu)=\frac{ K \sqrt{B}}{A\sqrt{2}}\tan\big(\sqrt{\frac{B}{2}}(\nu+\delta_0)\big)-\frac{BK}{2A}\ \nu+\Phi_0.
		\end{align}

\noindent \textbf{(viii)} \  If $B<0$, we obtain that
		\begin{align}
			&\Psi=-\varepsilon\sqrt{\frac{-B}{2A}}\tanh \Big[\sqrt{-\frac{B}{2}} (\nu+\delta_0)\Big], \\
			&\Phi=-\frac{K\sqrt{-B}}{A\sqrt{2}}\tanh \Big[\sqrt{-\frac{B}{2}} (\nu+\delta_0)\Big]-\frac{BK}{2A}\ \nu+\Phi_0.
		\end{align}

		\subsection{A lump type stationary solution }
		Following the work \cite{Ozawa}, we suggest a solution of the form
		\begin{equation}
			\label{lump} \psi (x,y,t)= \frac{1}{\Omega(x,y)}, \qquad \varphi(x,y,t)=\lambda_3 \frac{ \ \dy \Omega(x,y)}{\Omega(x,y)}
		\end{equation}
		  to \eqref{BRZRa}-\eqref{BRZRb}, where $\Omega(x,y)=\lambda_1 x^2 +\lambda_2 y^2 + 1$. Under some restrictions for $a$ and $b$, we determine $\lambda_1,\lambda_2, \lambda_3$ as
		\begin{equation}
			\lambda_1 =\frac{mw-b}{8a}, \qquad \lambda_2=\frac{b-mw}{8\epsilon_0},\qquad \lambda_3=\frac{4 \epsilon_0}{(mc-d)w}
		\end{equation}
		when
		\begin{equation}
			a=\frac{\epsilon_0}{mc^2 -1} ,\qquad b=\frac{(c^2 m^2-d^2 -2m +2mcd)w}{2(m c^2 -1)}.
		\end{equation}
The solution \eqref{lump} can be used as an initial condition to obtain the development of a lump profile with numerical schemes. We plot  this stationary solution for the values of the parameters  $d=\epsilon_0=w=1$ and $c=1/4$, $m=1/6$; that is, explicitly,
        \begin{equation}
			\label{lumpfig} \psi (x,y,t)= \frac{1}{1 + 0.0574 x^2 + 0.0580 y^2}, \qquad \varphi(x,y,t)=-\frac{0.4842 y}{1 + 0.0574 x^2 + 0.0580 y^2}
		\end{equation}
is pictured in Figure 3.		
		\begin{figure*}[h!]
			%%\begin{minipage"}{.5\linewidth}
			\centering
			\subfigure[]{\label{main:a}\includegraphics[scale=.50]{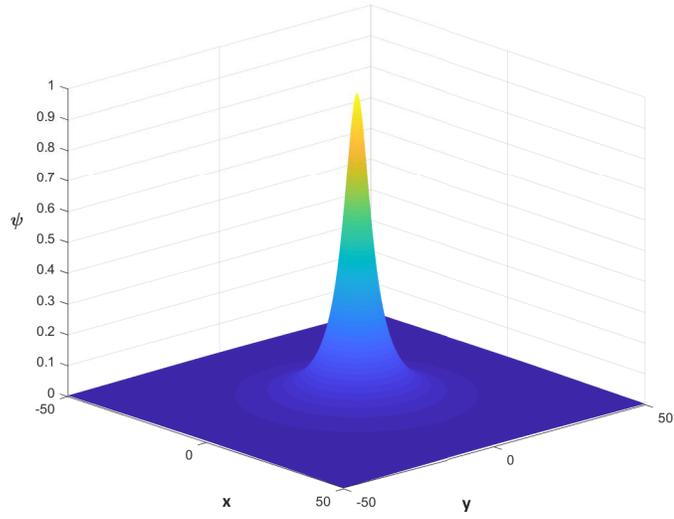}}
			%%\end{minipage}%
			%%\begin{minipage}{.5\linewidth}
			\centering
			\subfigure[]{\label{main:b}\includegraphics[scale=.50]{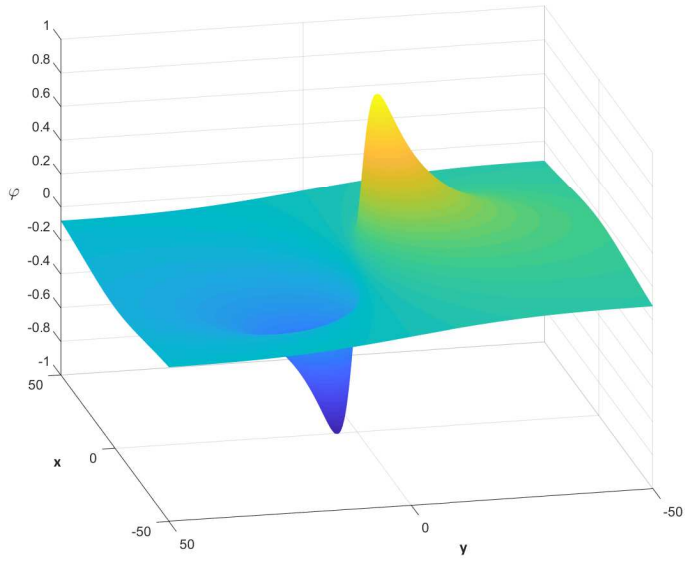}}
			%%\end{minipage}\par\medskip
			\caption{Plots of \eqref{lumpfig}. The values of parameters are $d=\epsilon_0=w=1$ and $c=1/4$, $m=1/6$.}
			\label{fig:main}
		\end{figure*}

\section{Conclusion}
In this article we focused on the Benney--Roskes/Zakharov--Rubenchik system of equations. The recent literature includes works which have taken into consideration
this system from several approaches. The system has connections with the well-known widely studied Davey--Stewartson system of equations. From the group-theoretical point of view, the DS system of equations
admits an infinite-dimensional Lie symmetry algebra of KMV type exactly in the integrable case, which is a feature shared by some other integrable equations in (2+1)-dimensions. Besides, for the DS system,  there is also a vast amount of literature
devoted to the search for exact solutions. The fact that the BR/ZR system includes the DS system in the limiting case motivated us to investigate the Lie invariance algebra of the BR/ZR system and also search for any
type of exact solutions that might contribute to the available literature.

Through our analysis we identified the Lie algebra of the BR/ZR system also as an infinite-dimensional one. We were also successful in obtaining exact solutions of several types, i.e. periodic, line soliton and stationary solutions. We believe these solutions will draw the attentions of the community and they might support and motivate further research on the BR/ZR system.

\section*{Acknowledgement}
We would like to thank Prof. Faruk Güngör for carefully reading the manuscript and for his valuable suggestions.

	\end{document}